\documentclass[a4paper,10pt,twoside]{cpc-hepnp}
\usepackage{CJK,upgreek,fancyhdr}
\usepackage{multicol}
\usepackage{graphicx}
\usepackage{booktabs}
\usepackage{amssymb,bm,mathrsfs,bbm,amscd}
\usepackage[tbtags]{amsmath}
\usepackage{lastpage}
\usepackage{color}

\begin{document}
\begin{CJK*}{GB}{gbsn}

\def\vsigma{{\hbox{\boldmath $\sigma$}}}
\def\bp{{\hbox{\boldmath $p$}}}
\def\bk{{\hbox{\boldmath $k$}}}
\def\bq{{\hbox{\boldmath $q$}}}
\def\undersim#1{\setbox9\hbox{${#1}$}{#1}\kern-\wd9\lower
    2.5pt \hbox{\lower\dp9\hbox to \wd9{\hss $_\sim$\hss}}}

\fancyhead[c]{\small Chinese Physics C~~~Vol. xx, No. x (201x) xxxxxx}
\fancyfoot[C]{\small xxxxxx-\thepage}

\footnotetext[0]{Received xx February 2017}

\title{Pion transverse-momentum spectrum, elliptic flow, and Hanbury-Brown-Twiss
interferometry in a viscous granular source model\thanks{Supported by National
Natural Science Foundation of China (11675034 and 11275037) }}

\author{%
      Jing Yang (Ñîæº)$^{1}$%
\quad Wei-Ning Zhang (ÕÅÎÀÄþ)$^{1,2;1)}$\email{wnzhang@dlut.edu.ac.cn}%
\quad Yan-Yu Ren (ÈÎÑÓÓî)$^{2}$
}
\maketitle

\address{%
$^1$ School of Physics and Optoelectronic Technology, Dalian University of
Technology, Dalian, Liaoning 116024, China\\
$^2$ Department of Physics, Harbin Institute of Technology, Harbin, Heilongjiang
150006, China\\
}

\begin{abstract}
We examine the evolution of quark-gluon plasma (QGP) droplets with viscous
hydrodynamics and analyze the pion transverse-momentum spectrum, elliptic flow,
and Hanbury-Brown-Twiss (HBT) interferometry in a granular source model
consisting of viscous QGP droplets.  The shear viscosity of the QGP
droplet speeds up and slows down the droplet evolution in the central and
peripheral regions of the droplet, respectively.  The effect of the bulk
viscosity on the evolution is negligible.  Although there are viscous effects
on the droplet evolution, the pion momentum spectrum and elliptic flow change
little for granular sources with and without viscosity.
On the other hand, the influence of viscosity on HBT radius $R_{\rm out}$
is considerable.  It makes $R_{\rm out}$ decrease in the granular source
model.  We determine the model parameters of granular sources using the
experimental data of pion transverse-momentum spectrum, elliptic flow,
and HBT radii together, and investigate the effects of viscosity on the
model parameters.  The results indicate that the granular source model
may reproduce the experimental data of pion transverse-momentum spectrum,
elliptic flow, and HBT radii in heavy-ion collisions of Au-Au at
$\sqrt{s_{NN}}=200$ GeV and Pb-Pb at $\sqrt{s_{NN}}=2.76$ TeV in different
centrality intervals.  The viscosity of the droplet leads to an increase in
the initial droplet radius and a decrease of the source shell parameter
in the granular source model.
\end{abstract}



\footnotetext[0]{\hspace*{-3mm}\raisebox{0.3ex}{$\scriptstyle\copyright$}2013
Chinese Physical Society and the Institute of High Energy Physics
of the Chinese Academy of Sciences and the Institute
of Modern Physics of the Chinese Academy of Sciences and IOP Publishing Ltd}%

\begin{multicols}{2}

\section{Introduction}

At sufficiently high temperature and/or high density, quantum chromodynamics
(QCD) predicts a transition from hadronic matter to the de-confined phase of
quarks and gluons, called Quark-Gluon Plasma (QGP) \cite{GroPisYaf81-rmp}.
It is the primary goal of ultrarelativistic heavy ion collisions to create
the QGP and study its evolution and detection
\cite{BRAHMS05,PHOBOS05,STAR05,PHENIX05}.
Relativistic hydrodynamics is an efficient tool for describing the evolution
of the hot and dense matter produced in high-energy heavy-ion collisions
\cite{{Ris98,KolHei03,HamKodSoc04,MurRis04,HeiSonCha06,BaiRomWie06,SonHei08,
SchenkeJeonGale10,MonHirano10,Romatschke10,Song11,Huovinen13,GaleJeonSchenke13,
JeonHeinz15,SouKoiKod16}}.
Ideal hydrodynamics, with the strong assumptions that the fluid evolves
without dissipation, has demonstrated that it can reproduce many
experimental results for heavy-ion collisions at the Relativistic
Heavy Ion Collider (RHIC) \cite{{BRAHMS05,PHOBOS05,STAR05,PHENIX05,Gyu04,
Huo&KolHei04,GyuMcL05,Shuryak05,MulNag06}}.
However, dissipative evolution of the hot and dense matter is a more general
case, and viscous hydrodynamic models have recently been applied in  heavy
ion collisions at the RHIC and the Large Hadron Collider (LHC)
\cite{{MurRis04,HeiSonCha06,BaiRomWie06,SonHei08,SchenkeJeonGale10,MonHirano10,
Romatschke10,Song11,Huovinen13,GaleJeonSchenke13,JeonHeinz15,SouKoiKod16,
Cha-prc06,Mur-prc07,Rom-PRL07,DusTea-prc08,LuzRom-prc08,LuzRom-PRL09,PraVre08,
Pratt09,MonHir-prc09,SonHei-prc10,Bozek11,BozekWysk12,s95p-Shen10,SchJeoGal-PRL11,
Song-prc11,RoyCha-prc12,DusSch-prc12,HeiSne13,Chaudhuri-ahep13,Vuj-prc14,Rugg14,
Song-prc11a,Shen-prc11,SchenkeJeonGale-plb11,ShenHeinz-prc12,Soltz13,Karpenko15}}.

Because of the complexity of the processes of high-energy heavy-ion collisions,
it is hard to obtain the final-state results directly from first-principle
calculations of QCD, the fundamental theory of strong interaction.
Phenomenologically, starting from the experimental data of final observables
one may infer the processes backward step-by-step with models.
In Refs. \cite{YangZhang14-ahep} and \cite{YangZhang15-Gyongyos},
we systematically investigated the pion transverse-momentum spectrum, elliptic
flow, and Hanbury-Brown-Twiss (HBT) interferometry in a granular source model
of QGP droplets and compared the model results of these observables with the
experimental data of heavy-ion collisions at the RHIC and LHC.  In this model,
granular sources are assumed to be formed at a later time in the QGP expansion.
The lumps of QGP after this time are dealt with as spherical droplets
for simplicity, which evolve in ideal hydrodynamics as in previous granular source
models \cite{{Zhang04-prc,Zhang06-prc,Zhang09-prc,Zhang11-cpl,Zhang11-Tokyo}}.
The anisotropic pressure gradient in the early QGP matter is assumed to lead to
the anisotropic initial velocities of the QGP droplets in the granular source model.
The investigations indicated \cite{YangZhang14-ahep} that the granular source
model consistently reproduces the data of pion transverse-momentum spectrum, elliptic flow,
and HBT radii in heavy ion collisions at the RHIC and the LHC.  On this basis, developing a viscous granular source model
and examining the effect of viscosity on the source evolution and the
multi-observable analyses in the viscous granular source model are the
motivations of this work.  Furthermore, experimental data for different
observables at different collision energies and in different centrality
intervals give strict constraints on the model parameters.  This makes these
parameters include useful information about the source geometry and dynamics.
They may provide helpful information for the study of the QGP properties and
dynamics at earlier stages.  Therefore, it is of interest to investigate the
applicability of the parameters and their modification in a viscous granular
source model.

In this paper, we use viscous relativistic hydrodynamics to describe
droplet evolution, and construct the granular source with the evolving
droplets.  We analyze the pion transverse-momentum spectrum, elliptic
flow, and HBT interferometry in the viscous granular source model.
The model parameters of granular sources are determined by the experimental
data of pion transverse-momentum spectrum, elliptic flow, and HBT radii
together.  The effects of viscosity on the model parameters are investigated.
We find that the motion of droplets in the granular sources reduces the
effect of viscosity on the pion transverse-momentum spectrum and the
viscosity causes a small decrease of the elliptic flow.  However,
viscosity makes the HBT radius $R_{\rm out}$ decrease significantly in
the granular source model.  The model results of the granular sources
are in accordance with the experimental data of  Au-Au collisions at
the RHIC and  Pb-Pb collisions at the LHC.  This means that the granular
source model reflects in some degree the physics of the system evolution
during the later stages of heavy-ion collisions.

The rest of this paper is organized as follows.  In Section 2, we review
the second order M\"{u}ller-Israel-Stewart formalism for viscous relativistic
hydrodynamics.  We also present the equations of the viscous hydrodynamics in
spherical geometry and provide their numerical solutions for the droplet.  In Section 3, we analyze the pion transverse-momentum
spectrum, elliptic flow, and HBT interferometry for viscous granular sources
for heavy-ion collisions of Au-Au at $\sqrt{s_{NN}} =200$ GeV and Pb-Pb
at $\sqrt{s_{NN}}=2.76$ TeV.  In Section 4, we compare the results of the
observables for granular sources with different model-parameter sets and
investigate the influence of viscosity on the parameters.  Finally, we provide
a summary and discussion in Section 5.

\section{Evolution of QGP droplets with viscous hydrodynamics}

In our granular source model, the source evolution is regarded as the
superposition of the evolutions of the QGP droplets, and the droplets are
assumed to have spherical geometry for simplicity
\cite{{YangZhang14-ahep,YangZhang15-Gyongyos,Zhang04-prc,Zhang06-prc,
Zhang09-prc,Zhang11-cpl,Zhang11-Tokyo}}.  In this section, we will present
briefly the description of viscous hydrodynamics for spherical QGP
droplets, and show their hydrodynamical evolution.

\subsection{Viscous hydrodynamics for spherical QGP droplets}
For the zero-net-baryon systems produced in ultrarelativistic heavy ion collisions,
the description of viscous hydrodynamics starts from the local conservation of
energy and momentum,
\begin{equation}
d_{\mu}T^{\mu\nu}~{\equiv}~
\partial_{\mu}T^{\mu \nu}+\Gamma^{\mu}_{\mu \lambda}T^{\lambda \nu}
+T^{\mu\lambda}\Gamma^{\nu}_{\lambda \mu}=0.
\label{conservationlaw}
\end{equation}
Here, $d_\mu$ denotes the covariant differential, and the Christoffel
symbol is given by
\begin{equation}
\Gamma^{\gamma}_{\alpha \beta}~{\equiv}~\frac{1}{2}g^{\gamma \sigma}
\left({\partial_\alpha g_{\beta\sigma}+\partial_\beta g_{\sigma \alpha}
-\partial_\sigma g_{\alpha \beta}}\right),
\end{equation}
where $g^{\mu\nu}=\text{diag}(+,-,-,-)$ is a metric tensor.  We adopt the
Landau-Lifshitz frame \cite{Landau-Lifshitz}, i.e., the local rest frame
of energy density.  The energy-momentum tensor $T^{\mu\nu}$ is
\begin{equation}
T^{\mu\nu}=\epsilon u^\mu u^\nu-(P+\Pi)\Delta^{\mu\nu}+\pi^{\mu\nu},
\end{equation}
where $\epsilon$ is the local energy density, $u^\mu$ is the four-velocity
of the energy flow, $P$ and $\Pi$ are the thermal equilibrium pressure and
the bulk viscous pressure respectively, $\Delta^{\mu\nu}~{\equiv}~g^{\mu\nu}-u^\mu u^\nu$,
and $\pi^{\mu\nu}$ is the shear viscous pressure tensor.

Based on the second-order M\"{u}ller-Israel-Stewart theory of dissipative
hydrodynamics \cite{2ndMIS}, $\pi^{\mu\nu}$ and $\Pi$ satisfy the viscous relaxation
equations
\begin{equation}
\begin{aligned}
\tau_{\pi}\Delta^{\mu\alpha}\Delta^{\nu\beta}
&D\pi_{\alpha\beta}+\pi^{\mu\nu}\\
=&2\,\tilde{\eta}\,\nabla^{<\mu}u^{\nu>} \!-\frac{1}{2}\pi^{\mu\nu}\,
\tilde{\eta}\,T d_\lambda   \left({\frac{\tau_\pi u^\lambda}{\tilde{\eta}T}}
\right)
\end{aligned}
\label{2ndMIS-1}
\end{equation}
and
\begin{equation}
\begin{aligned}
\tau_{\Pi}D\Pi+\Pi
=-\tilde{\xi}\,d_{\mu}u^{\mu}-\frac{1}{2}\Pi\,\tilde{\xi}\,T d_\lambda
  \left({\frac{\tau_\pi u^\lambda}{\tilde{\xi}T}}\right),
\end{aligned}
\label{2ndMIS-2}
\end{equation}
respectively.  Here, $\tilde{\eta}$ is the shear viscosity coefficient,
$\tilde{\xi}$ is the bulk viscosity coefficient, $\tau_\pi$ is the relaxation
time for shear viscosity, and $\tau_\Pi$ is the relaxation time for bulk
viscosity.  In our numerical calculations, quantities $\tilde{\eta}$,
$\tilde{\xi}$, $\tau_\pi$, and $\tau_\Pi$ are taken to have certain
relationships with entropy density which will be discussed later.
Some notations used in Eqs. (\ref{2ndMIS-1}) and (\ref{2ndMIS-2}) are
defined as
\begin{equation}
\begin{aligned}
&d_\mu u^\nu~{\equiv}~\partial_\mu u^\nu+\Gamma_{\alpha \mu}^{\nu}u^\alpha,\\
&D~{\equiv}~u^\mu d_\mu,~~\nabla^\mu \equiv \Delta^{\mu \nu} d_\nu,\\
&\nabla^{<\mu}u^{\nu>} \equiv \nabla^{(\mu}u^{\nu)}-\frac{1}{3}
\Delta^{\mu\nu}d_\lambda u^{\lambda},\\
&A^{(\alpha\beta)}~{\equiv}~\frac{1}{2}\left({A^{\alpha\beta}+A^{\beta\alpha}}
\right).
\end{aligned}
\end{equation}

For a spherical droplet, Eqs. (\ref{conservationlaw}), (\ref{2ndMIS-1}),
and (\ref{2ndMIS-2}) can be written as follows (see Appendix A for
detailed derivations), which are suitable for solving numerically:
\end{multicols}
\ruleup
\begin{equation}
\left\{{~
\begin{aligned}
&\frac{\partial}{\partial t}\,E+\frac{\partial}{\partial r}\,[(E+P_r)v_r]
=-(E+P_r)\,\frac{2v_r}{r},\\
&\frac{\partial}{\partial t}\,M+\frac{\partial}{\partial r}\,[Mv_r+P_r]
=-M\,\frac{2v_r}{r}-\frac{2\tau^{rr}-\tau^{\theta\theta}-\tau^{\phi\phi}}{r},\\
&\frac{\partial}{\partial t}\,\tau^{rr}+\frac{\partial}{\partial r}\,[v_r
\tau^{rr}]
=\sigma^{rr}\left({2\frac{1}{\gamma}~\frac{\tilde{\eta}}{\tau_\pi}}\right)
+\tau^{rr}\left[{-\frac{1}{\gamma}\,\frac{1}{\tau_\pi}-\frac{1}{2}\,
\frac{1}{\gamma}\,\Theta-\frac{1}{2}\,\frac{1}{\gamma}\,\frac{\tilde{\eta}T}
{\tau_\pi}D\left({\frac{\tau_\pi}{\tilde{\eta}T}}\right)+\frac{\partial v_r}
{\partial r}}\right],\\
&\frac{\partial}{\partial t}\,\tau^{\theta\theta}+\frac{\partial}{\partial r}\,
[v_r\tau^{\theta\theta}]
=\sigma^{\theta\theta}\left({2\frac{1}{\gamma}~\frac{\tilde{\eta}}{\tau_\pi}}
\right)+\tau^{\theta\theta}\left[{-\frac{1}{\gamma}\,\frac{1}{\tau_\pi}-
\frac{1}{2}\,\frac{1}{\gamma}\,\Theta-\frac{1}{2}\,\frac{1}{\gamma}\,
\frac{\tilde{\eta}T}{\tau_\pi}D\left({\frac{\tau_\pi}{\tilde{\eta}T}}\right)
+\frac{\partial v_r}{\partial r}}\right],\\
&\frac{\partial}{\partial t}\,\tau^{\phi\phi}+\frac{\partial}{\partial r}\,
[v_r\tau^{\phi\phi}]
=\sigma^{\phi\phi}\left({2\frac{1}{\gamma}\,\frac{\tilde{\eta}}{\tau_\pi}}
\right)+\tau^{\phi\phi}\left[{-\frac{1}{\gamma}\,\frac{1}{\tau_\pi}
-\frac{1}{2}\,\frac{1}{\gamma}\,\Theta-\frac{1}{2}\,\frac{1}{\gamma}\,
\frac{\tilde{\eta}T}{\tau_\pi}D\left({\frac{\tau_\pi}{\tilde{\eta}T}}\right)
+\frac{\partial v_r}{\partial r}}\right],\\
&\frac{\partial}{\partial t}\,\Pi+\frac{\partial}{\partial r}\,[v_r\Pi\,]
=\Theta\,\frac{1}{\gamma}\,\frac{\tilde{\xi}}{\tau_\pi}
+\Pi\left[{-\frac{1}{\gamma}\,\frac{1}{\tau_\pi}-\frac{1}{2}\,\frac{1}{\gamma}
\,\Theta-\frac{1}{2}\,\frac{1}{\gamma}\,\frac{\tilde{\xi}T}{\tau_\pi}D\left(
{\frac{\tau_\pi}{\tilde{\xi}T}}\right)+\frac{\partial v_r}{\partial r}}\right],
\end{aligned}
}\right.
\label{single-evol}
\end{equation}
\ruledown \vspace{0.5cm}
\begin{multicols}{2}
\hspace*{-5mm}where
\begin{eqnarray}
&E=\gamma^2 (\epsilon+v_r^2 P_r), \\
&P_r=P+\Pi+\tau^{rr}, \\
&M=v_r(E+P_r),
\end{eqnarray}
and $\tau^{rr}$, $\tau^{\theta\theta}$, and $\tau^{\phi\phi}$ are three
quantities introduced in the nonzero components of $\pi^{\mu\nu}$ as
\begin{equation}
\pi^{\mu\nu}
=\left(
\begin{array}{cccc}
\gamma^2 v_r^2\tau^{rr}&\gamma^2 v_r \tau^{rr}&0                         &0\\
\gamma^2 v_r\tau^{rr}  &\gamma^2 \tau^{rr}    &0                         &0\\
0                      &0                     &r^{-2}\tau^{\theta\theta} &0\\
0                      &0                     &0                         & (r\sin\theta)^{-2}\tau^{\phi\phi}\\
\end{array}
\right),
\end{equation}
where
\begin{equation}
\tau^{rr}+\tau^{\theta\theta}+\tau^{\phi\phi}=0,
\label{tau=0}
\end{equation}
from the traceless condition $\pi^\mu_{~\mu}=0$.  The other referred quantities
in Eq. (\ref{single-evol}) are given by
\begin{eqnarray}
&&\Theta
=\frac{\partial\gamma}{\partial t}+\frac{\partial(\gamma v_r)}{\partial r}
 +2\gamma\frac{v_r}{r},\\
&&\sigma^{rr}=-2\sigma^{\theta\theta}=-2\sigma^{\phi\phi}
=2\left({\gamma\frac{v_r}{r}-\frac{1}{3}\Theta}\right).
\label{sigma}
\end{eqnarray}

\subsection{Hydrodynamical evolution of QGP droplets}
In equation set (\ref{single-evol}), there are eight variable quantities, $E$ (or $\epsilon$), $P_r$
(or $P$), $M$ (or $v_r$), $\tau^{rr}$, $\tau^{\theta\theta}$, $\tau^{\phi\phi}$,
$\Pi$, and $T$.  With the traceless condition
[Eq. (\ref{tau=0})], there are seven independent variables.  An equation
of state (EOS), $P(\epsilon,T)$, is needed when solving the hydrodynamic
equations.  In the numerical calculations in this paper, we use the EOS
s95p-PCE \cite{s95p-Shen10}, which combines the lattice QCD data at high
temperature and the hadron resonance gas at low temperature.  The equations
in (\ref{single-evol}) have a similar form, $\partial_t U+\partial_r F(U)
=G(U)$.  They can be solved by the HLLE algorithm
\cite{Ris98,Ris9596,Zhang04-prc,ZhaEfa,HuZha15}.

The initial energy density distribution of a droplet in our granular source model
is taken to be the Woods-Saxon form as in Ref. \cite{YangZhang15-Gyongyos},
\begin{equation}
\epsilon(r)=\epsilon_0\frac{1}{e^{\frac{r-r_0}{a}}+1},
\label{Eq-woodssaxon}
\end{equation}
where $\epsilon_0$ is the initial energy density at the center of the droplet,
$r_0$ denotes the initial droplet radius, and $a$ is the Woods-Saxon width parameter,
which is taken to be $0.1r_0$.  The shear and bulk viscosity quantities,
$\tau^{rr}$, $\tau^{\theta\theta}$, $\tau^{\phi\phi}$, and $\Pi$ are set to be
zero initially because the droplet is initially at rest  in the local frame.

When solving equation set (\ref{single-evol}) numerically, we also need to
specify the viscosity coefficients $\tilde{\eta}$ and $\tilde{\xi}$ and the
relaxation time $\tau_{\pi}$ and $\tau_{\Pi}$.
For the strongly coupled QGP matter produced in ultrarelativistic heavy ion
collisions, the ratio of the shear viscosity to entropy density, $\tilde{\eta}
/s$, is expected to be between 0.08--0.24 \cite{LuzRom-prc08,Song11}, that is 1 to 3
times the minimum $(\tilde{\eta}/s)_{\rm min}=1/4\pi$ \cite{KovSonSta-PRL05}.
To look into the effects of viscosity on observables, we take the upper bound
of $\tilde{\eta}/s$ (3 times the minimum) in calculations.
The shear relaxation time $\tau_\pi$ is taken as $3\tilde{\eta}/(sT)$ as in
Refs. \cite{SonHei-prc10,Song11,Bozek11,BozekWysk12,s95p-Shen10}.
For bulk viscosity, the minimum $(\tilde{\xi}/s)_{\rm min}$ obtained by the
AdS/CFT result is $2(\tilde{\eta}/s)_{\rm min}(1/3-c_s^2)$ \cite{AdSCFT-bulk}.
Here, $c_s$ is the speed of sound.
The entropy density $s$ and the speed of sound  $c_s$ as functions of energy
density or temperature are given by the EOS s95p-PCE (see Fig. 2 and Eq.
(A2) in Ref. \cite{s95p-Shen10}).  We take $\tilde{\xi}/s$ also as 3 times
the minimum $(\tilde{\xi}/s)_{\rm min}$ in calculations.  The bulk relaxation
time is taken to be the parametrization, $\tau_\Pi=\max[120 \cdot\tilde{\xi}
/s,0.1]~\text{fm}/c$, as in Ref. \cite{SonHei-prc10}.

\begin{center}
\vspace*{-3mm}
\includegraphics[width=7cm]{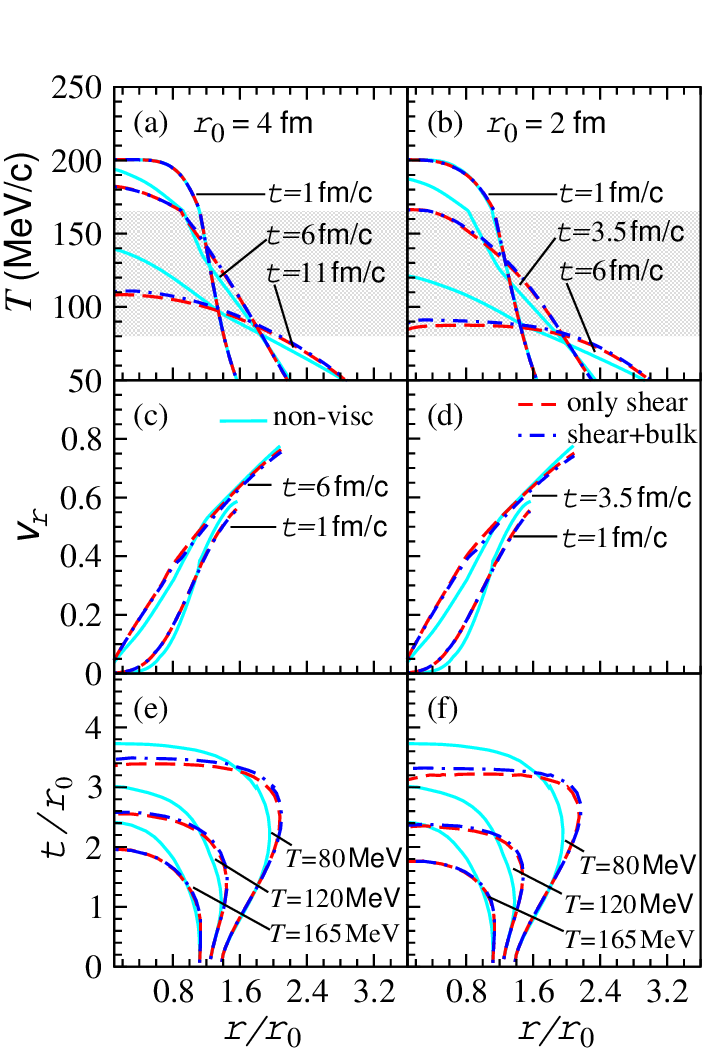}
\figcaption{\label{Fig-drop-evo}   (Color online) Temperature evolution [(a)
and (b)], expansion velocity [(c) and (d)], and isothermals [(e) and (f)] of
droplets with different initial radii and initial energy density
$\epsilon_0=2.2$ GeV/fm$^3$. }
\end{center}

We plot in Fig. \ref{Fig-drop-evo} the temperature evolution (top panels),
expanding velocity (middle panels), and isothermals (bottom panels) of the
droplet with initial radii $r_0=$4 and 2 fm.
Here the cyan solid lines, red dashed lines, and blue dot-dashed lines
are for droplets without viscosity, with only shear viscosity, and with
both shear and bulk viscosities.  The initial energy density is taken as
$\epsilon_0=2.2$ GeV/fm$^3$ \cite{YangZhang14-ahep,YangZhang15-Gyongyos},
corresponding to the moment of granular source formation.  Also, the marked
time in Fig. \ref{Fig-drop-evo} is evolution time relative to the formation
time.  The shear viscosity of the droplet has a dual effect on the droplet
evolution in the edge and central regions of the droplet.  In the edge region,
the large pressure gradient leads to a fast expansion and the viscosity plays
to resist the expansion.  So, the temperature of a viscous droplet decreases
a little slower than that of a droplet without viscosity.
However, the expansion velocity is small in the central region of the droplet
because of small pressure gradient.  In this case, the viscosity pulls the
matter in the central region to move and leads to a lower temperature of
the viscous droplet compared to that of the droplet without viscosity in the
central region.  The dual effects on the droplet expansion velocity and
isothermals can also be observed in the middle and bottom panels of Fig.
\ref{Fig-drop-evo}.  By comparing the curves, considering only shear
viscosity and both shear and bulk viscosities, one can see that the influence
of bulk viscosity on the evolution is very small.

In our granular source model, the particles are emitted from the evolving
droplets.  To investigate the pion transverse-momentum $p_T$ spectrum of
the granular source, we first examine the $p_T$ spectrum of pions emitted from a
single droplet.  The pion transverse-momentum spectrum for the evolving
single droplet are calculated with the Cooper-Frye formula
\cite{Cooper-Frye}
\begin{equation}
\frac{1}{2\pi}\frac{d^2N}{p_Tdp_Tdy}\propto \frac{1}{(2\pi)^3} \int_{\Sigma}
p^{\mu} d\sigma_{\mu} (f_0+\delta f),
\label{Cooper-Frye}
\end{equation}
where $y$ is particle rapidity, $f_0$ is the Bose-Einstein distribution, and
$\delta f$ is given by \cite{DusTea-prc08,SonHei08,s95p-Shen10},
\begin{equation}
\delta f=f_0(1+f_0)\frac{p^\mu p^\nu}{2(\epsilon+P)T^2}
(\pi_{\mu\nu}-\frac{2}{5}\Pi\Delta_{\mu\nu}).
\end{equation}
The integration in Eq. (\ref{Cooper-Frye}) is over the hypersurface at
freeze-out temperature $T_f$.

In Fig. \ref{Fig-drop-spe}, we plot the transverse-momentum spectra of pions
emitted from droplets evolving as shown in Fig. \ref{Fig-drop-evo} in the
freeze-out temperature region $80<T_f<T_c$ ($T_c=165$ MeV), with the following
probability used in the granular source models \cite{YangZhang14-ahep,YangZhang15-Gyongyos,Zhang09-prc,Zhang11-cpl,Zhang11-Tokyo},
\begin{eqnarray}
\frac{dP}{dT_f} \,{\propto}\, f_{\text{dir}}\,e^{-\frac{T_c-T_f}
{\Delta T_{\text{dir}}}}+({1-f_{\text{dir}}})
e^{-\frac{T_c-T_f}{\Delta T_{\text{dec}}}}.
\label{dPdTf}
\end{eqnarray}
Here, $f_{\text{dir}}$ is the fraction of direct emission around $T_c$, and
$\Delta T_{\text{dir}}$ and $\Delta T_{\text{dec}}$ are the temperature widths
for the direct and decay emissions, respectively.  In the calculations, we take
$f_{\text{dir}}=0.75$, $\Delta T_{\text{dir}}=10$ MeV, and $\Delta T_{\text{dec}}
=90$ MeV as in \cite{YangZhang14-ahep,YangZhang15-Gyongyos}.
In the edge region of the droplet, the large expansion velocity causes the
large average  transverse momentum of the particles.  The viscous droplet in the edge
region evolves more slowly (thus has higher temperature) than the droplet
without viscosity.  So, the transverse momentum spectrum for viscous droplets
is higher at large $p_T$ than that for droplets without viscosity.
This effect is more obvious for smaller droplets.  The pressure gradient in
smaller droplets is larger than in larger droplets with the
same initial energy density.

\begin{center}
\includegraphics[width=7cm]{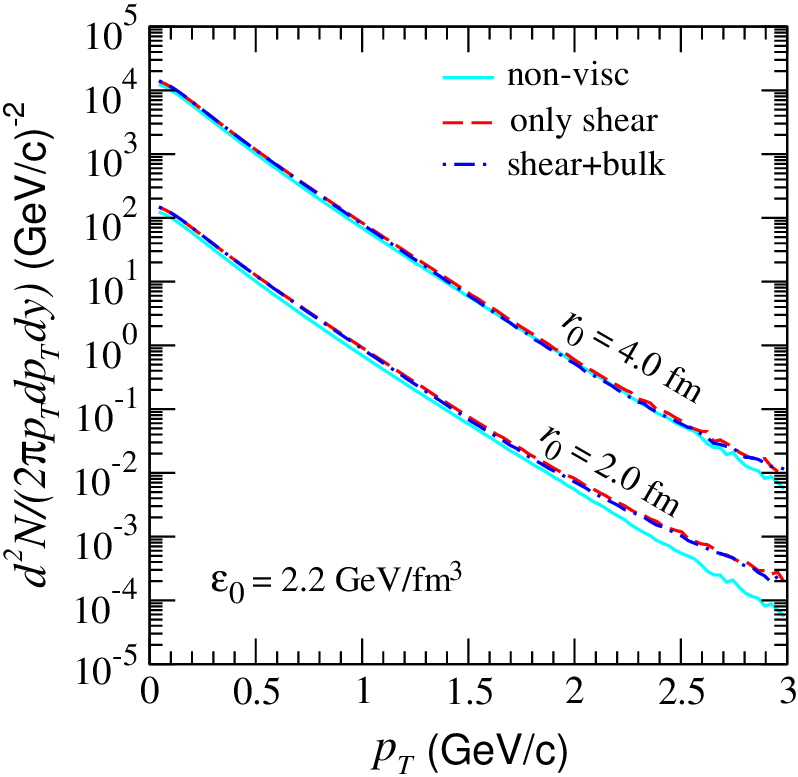}
\figcaption{\label{Fig-drop-spe}   (Color online) Transverse-momentum spectra
of pions emitted from evolving droplets as in Fig. \ref{Fig-drop-evo}, and
with the freeze-out temperature distribution (\ref{dPdTf}). }
\end{center}

\section{Multi-observable analyses for granular sources}
Single-particle momentum spectra, elliptic flow, and two-particle HBT
correlations are important observables in high-energy heavy-ion collisions.
They are tightly related to the initial conditions, evolution, and freeze-out of the particle-emitting sources.  In this section, we will analyze
these observables for the granular sources for  heavy ion collisions
at RHIC and LHC energies, and investigate the effects of viscosity on the
analysis results.

\subsection{Ingredients of granular source model}
In Refs. \cite{Zhang04-prc,Zhang06-prc,Zhang09-prc}, a granular source model
of QGP droplets was proposed and developed by Wei-Ning Zhang {\it et al.}
to explain the RHIC HBT puzzle, $R_{\text{out}}/R_{\text{side}}~\sim~1$
\cite{STAR-HBT-PRL01,PHENIX-HBT-PRL02,PHENIX-HBT-PRL04,STAR-HBT-prc05},
where $R_{\text{out}}$ and $R_{\text{side}}$ are two HBT radii in the transverse
plane along and perpendicular to the transverse momentum of the particle
pairs \cite{Bertsch88,Pratt90}.  Although the early idea of constructing
the granular source model was based on the first-order QCD transition
\cite{Zhang04-prc,Witten84,CseKap92,VenVis94,Randrup04}, the occurrence
of granular sources may not be limited to the first-order phase transition.
There are other factors, such as large initial fluctuations and some system
instabilities may lead to the granular inhomogeneous structure of
particle-emitting sources in ultrarelativistic heavy-ion collisions
\cite{{Zhang06-prc,Zhang09-prc,Zhang11-cpl,Zhang11-Tokyo,HuZha15,RenZha08-plb,
YanZha09-jpg,TorTomMis08-prc,TakTavQia09-PRL,WerKarPie10-prc,SchJeoGal11-PRL,
ISFFSC12}}.  In the granular source model, the droplets are assumed to have
spherical geometry and anisotropic initial velocities.
This model may be considered as a simplified picture of inhomogeneous
particle-emitting sources.

Apart from using viscous hydrodynamics to describing the droplet evolution,
we adopt all the ingredients of the granular source model used in Ref.
\cite{YangZhang15-Gyongyos}.  The initial energy density distribution
of a single droplet is assumed to be a Woods-Saxon distribution
\cite{YangZhang15-Gyongyos}, and the QGP droplets are initially distributed
within a cylinder along the beam direction ($z$-axis) by
\cite{YangZhang14-ahep,YangZhang15-Gyongyos}
\begin{eqnarray}
\frac{dN_d}{dx_0dy_0dz_0}\!\! &{\propto}&\!\!\Big[1-e^{-(x_0^2+y_0^2)/
\Delta {\cal R}_T^2}\Big]\theta({\cal R}_T- \rho_0)\cr
&&\times \theta({\cal R}_z -|z_0|),
\end{eqnarray}
where $\rho_0=\sqrt{x_0^2+y_0^2}$ and $z_0$ are the initial transverse
and longitudinal coordinates of the droplet centers, ${\cal R}_T$ and
${\cal R}_z$ describe the initial transverse and longitudinal sizes of
the source, and $\Delta {\cal R}_T$ is a transverse shell parameter.
Quantity $\Delta {\cal R}_T$ is used to describe the effect where
blast-wave-type expansion in early high density QGP matter may lead
to a void in the system central region \cite{Ris9596,Ris98} at the time
of the granular source formation ($\Delta {\cal R}_T\to 0$ for the uniform
transverse distribution of droplets).
The initial radius of the droplets, $r_0$, satisfies a Gaussian distribution
with standard deviation $\sigma_d$ in the droplet local frame.
We take the initial velocities of the droplets in the granular source as
\cite{YangZhang14-ahep,YangZhang15-Gyongyos}
\begin{equation}
\label{ini_v}
v_{{\rm d}i}=\mathrm{sign}(r_{0i}) \cdot a_i \bigg(\frac{|r_{0i}|}
{{\cal R}_i}\bigg)^{b_i},~~~~~~i=1,\,2,\,3,
\end{equation}
where $r_{0i}$ is $x_0$, $y_0$, or $z_0$ for $i=$ 1, 2, or 3, and
$\text{sign}(r_{0i})$ denotes the signal of $r_{0i}$, which ensures an
outward droplet velocity.  In Eq. (\ref{ini_v}), ${\cal R}_i=({\cal R}_T,
{\cal R}_T,{\cal R}_z)$, $a_i=(a_x,a_y,a_z)$ and $b_i=(b_x,b_y,b_z)$ are
the magnitude and exponent parameters in $x$, $y$, and $z$ directions respectively.
It is also convenient to use the equivalent parameters $\overline{a}_T =
(a_x+a_y)/2$ and $\Delta a_T=a_x-a_y$ instead of $a_x$ and $a_y$.  The
parameters $\overline{a}_T$ and $\Delta a_T$ describe the initial transverse
expansion and asymmetric dynamical behavior of the system, respectively.
For simplicity, we take $b_x=b_y=b_T$ in calculations.  The parameters
$b_T$ and $b_z$ describe the coordinate dependence of exponential power
in transverse and longitudinal directions.
The parameters ($\overline{a}_T$, $\Delta a_T$, $a_z$, $b_T$, $b_z$) reflect
the anisotropic coordinate-dependence of the droplet velocity caused by the
anisotropic pressure gradient in early QGP matter in the granular source
model.

\subsection{Pion momentum spectrum and elliptic flow in viscous granular
source model}
In high-energy heavy-ion collisions, the invariant momentum distribution
of final particles can be written in the form of a Fourier series
\cite{SV-YZ96,AP-SV98},
\begin{eqnarray}
\label{pdis}
E\frac{d^3N}{d^3p}=\frac{1}{2\pi}\frac{d^2N}{p_Tdp_Tdy}\left[1+\sum_n{2v_n
\cos(n\phi)}\right],
\end{eqnarray}
where $\phi$ is the particle azimuthal angle with respect to the reaction
plane.  In Eq. (\ref{pdis}), the first term on the right is the transverse
momentum spectrum in the rapidity interval $dy$, and the second harmonic
coefficient $v_2$ in the summation is called the elliptic flow.

In Fig. \ref{Figgs-spe}, we plot the pion transverse-momentum spectra
of granular sources without viscosity (cyan solid lines), with only
shear viscosity (red dashed lines), and with both shear and bulk viscosities
(blue dot-dashed lines), for Au-Au collisions at $\sqrt{s_{NN}}=200$ GeV
in the centrality intervals 0-5\%, 10-20\%, and 30-50\%, and for Pb-Pb
collisions at $\sqrt{s_{NN}}=2.76$ TeV in the centrality intervals 10-20\%
and 40-50\%.  In these centrality intervals, the experimental data of pion
transverse-momentum spectrum, elliptic flow, and HBT interferometry are all
available.  The experimental data plotted in Fig.
\ref{Figgs-spe} are from the PHENIX and STAR collaborations
at the RHIC \cite{PHE-spe04z,STA-spe04z} and the ALICE collaboration at
the LHC \cite{ALI-spe13z}.  In the calculations of the momentum spectra of
the granular sources, we take the rapidity cuts $|y|<0.1$ and $|y|<0.5$ to be
the same as in the experimental measurements at the RHIC \cite{STA-spe04z}
and at the LHC \cite{ALI-spe13z}, respectively.  The granular source
parameters are the same as in Ref. \cite{YangZhang15-Gyongyos} for the
Woods-Saxon initial energy distribution of droplets (see Set I of Table \ref{Tab-parameter}).

\begin{center}
\vspace*{2mm}
\includegraphics[width=7cm]{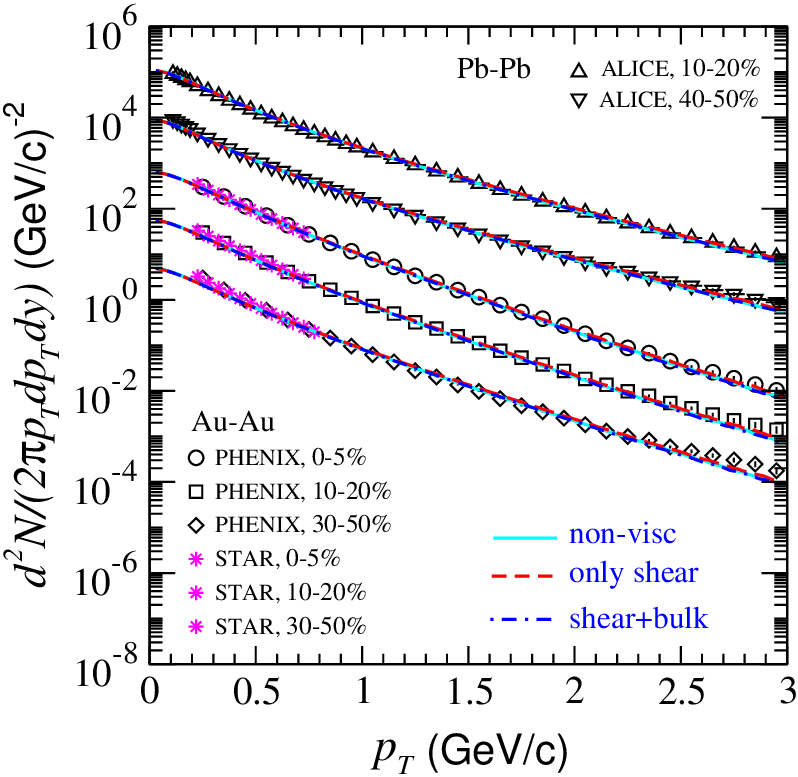}
\vspace*{2mm}
\figcaption{\label{Figgs-spe}   (Color online)
Pion transverse momentum spectra of granular sources without
viscosity (cyan solid lines), with only shear viscosity (red dashed
lines), and with both shear and bulk viscosities (blue dot-dashed
lines), for Au-Au collisions at $\sqrt{s_{NN}}=200$ GeV and Pb-Pb collisions at $\sqrt{s_{NN}}=2.76$ TeV.  The experimental data
measured by the PHENIX and STAR collaborations at the RHIC
\cite{PHE-spe04z,STA-spe04z} and by the ALICE collaboration
at the LHC \cite{ALI-spe13z} are also plotted. }
\end{center}

One can see from Fig. \ref{Figgs-spe} that the effect of viscosity on
the momentum spectrum is very small.  In the granular source model,
the momentum spectrum depends not only on the droplet evolution velocity,
which is viscosity related, but also on the outward motion of the droplet
as a whole, which is determined by the velocity parameters $a_T$ and $b_T$.
The outward motion of whole droplet boosts the particles emitted from
the droplet.  As a result, some particle momenta are decreased and some
particle momenta are increased compared to the case where the droplets
are at rest in the granular source ($a_T=0, a_z=0$).
Although viscosity may lead to a small enhancement of the
transverse-momentum distribution of a single droplet at larger $p_T$ (see
Fig. \ref{Fig-drop-spe}), the outward motion of the whole droplet lets some of
the particle contributions in the high transverse momentum region of the
single droplet spectrum move to the low transverse momentum region.
Meanwhile, it lets some of the particle contributions in the low transverse
momentum region of the single droplet spectrum move to the high transverse
momentum region.  As a result, the outward motion of the whole droplet partially reduces
 the influence of droplet viscosity on the momentum spectrum
in the granular source model.

\begin{center}
\vspace*{3mm}
\includegraphics[width=8cm]{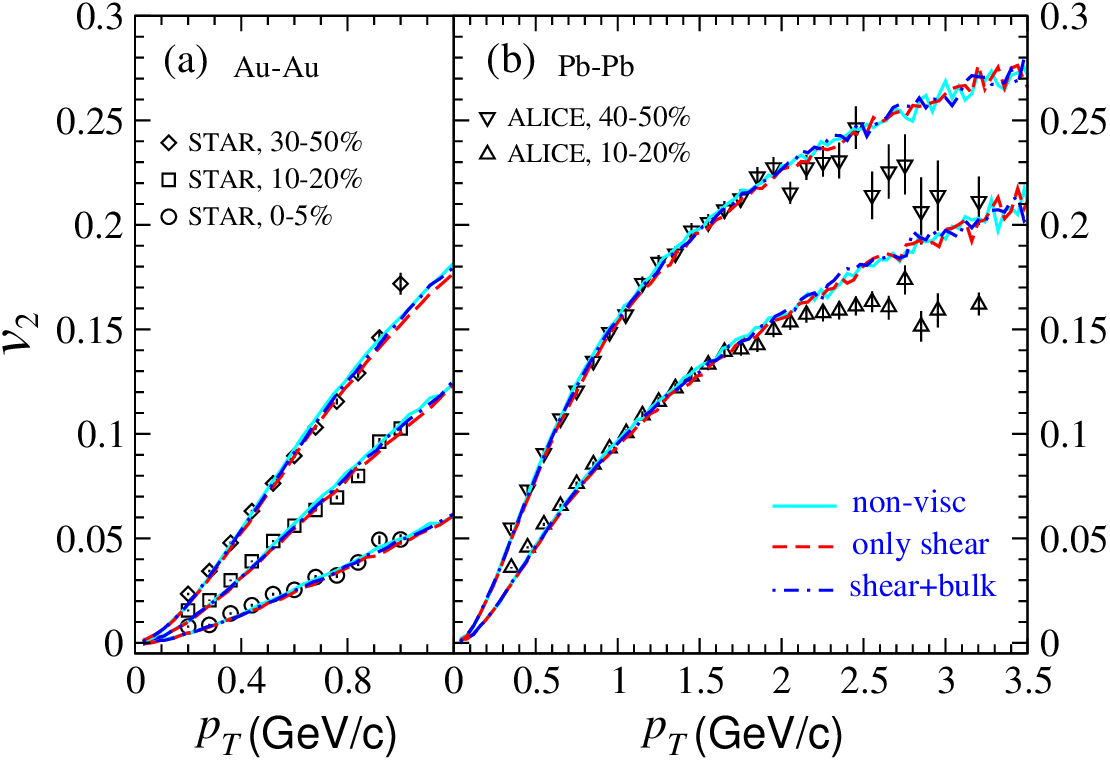}
\vspace*{3mm}
\figcaption{\label{Figgs-v2}   (Color online)
Pion elliptic flow of granular sources without viscosity
(cyan solid lines), with only shear viscosity (red dashed lines), and
with both shear and bulk viscosities (blue dot-dashed lines), for
Au-Au collisions at $\sqrt{s_{NN}}=200$ GeV [panel (a)] and Pb-Pb
collisions at $\sqrt{s_{NN}}=2.76$ TeV [panel (b)], respectively.
The experimental data measured by the STAR collaboration at the RHIC
\cite{STA-v2-05z} and by the ALICE collaboration at the LHC
\cite{ALI-v2-11z,ALI-v2-13z} are plotted in panels (a) and (b),
respectively. }
\end{center}

In Figs. \ref{Figgs-v2}(a) and \ref{Figgs-v2}(b), we show the results
of pion elliptic flow of the granular sources without viscosity (cyan solid
lines), with only shear viscosity (red dashed lines), and with both shear
and bulk viscosities (blue dot-dashed lines) as in Fig. \ref{Figgs-spe},
for Au-Au collisions at $\sqrt{s_{NN}}=200$ GeV and Pb-Pb collisions
at $\sqrt{s_{NN}}=2.76$ TeV, respectively.  The experimental data of elliptic
flow for the Au-Au and Pb-Pb collisions in the same centrality intervals as
the momentum-spectrum data in Fig. \ref{Figgs-spe}, measured by the
PHENIX and STAR collaborations at the RHIC \cite{PHE-spe04z,STA-spe04z} and
the ALICE collaboration at the LHC \cite{ALI-spe13z} are also shown in Figs.~\ref{Figgs-v2}(a) and \ref{Figgs-v2}(b), respectively.  We take the
pseudorapidity cuts $|\eta|<1.0$ and $|\eta|<0.8$ in the calculations of
$v_2$ of the granular sources to be the same as in the experimental analyses
carried out by the STAR \cite{STA-v2-05z} and the ALICE
\cite{ALI-v2-11z,ALI-v2-13z} collaborations, respectively.
The elliptic flow of the viscous granular source is
a little smaller than that of the granular source without viscosity.
Because the evolution of the spherical droplet is isotropic, the anisotropic
flow $v_2$ of the granular source is determined by the velocity parameters
$\Delta a_T$, $b_T$, and $b_z$ \cite{Zhang06-prc}.
These parameters reflect the anisotropic coordinate-dependence of the droplet
velocity caused by the anisotropic pressure gradient in the early QGP matter,
and are seldom affected by the droplet evolution afterwards in the granular source
model considered.
In Ref. \cite{YangZhang16}, the effects of droplet absorption of particles
on the momentum spectrum and HBT radii are investigated.
Further investigations of the influences of source evolution and absorption
on $v_2$ in the viscous granular source model will be of interest.
At larger $p_T$ ($>1.8$ GeV/$c$), the $v_2$ results of the granular sources
are larger than the experimental data.  This is because the current granular
source model does not include the hard processes which exist in high-energy
heavy-ion collisions.

\subsection{Pion interferometry results of granular sources}
The two-pion HBT correlation function is defined as the ratio of the two-particle
momentum spectrum $P(\bp_1,\bp_2)$ of identical pions to the product of the
two single-pion momentum spectra $P(\bp_1)P(\bp_2)$.  In interferometry
analyses in high-energy heavy-ion collisions, the two-pion correlation function
is usually fitted by the Gaussian parameterized formula
\begin{equation}
C(q_{\rm out},q_{\rm side},q_{\rm long})\!=\!1 \!+ \lambda\, e^{-R_{\rm out}^2
q_{\rm out}^2 -R_{\rm side}^2 q_{\rm side}^2 -R_{\rm long}^2 q_{\rm long}^2},
\label{Eq-CF}
\end{equation}
where $q_{\rm out}$, $q_{\rm side}$, and $q_{\rm long}$ are the Bertsch-Pratt
variables \cite{Bertsch88,Pratt90}, which denote the components of the relative
momentum $\bq=\bp_1-\bp_2$ in the transverse ``out" (parallel to the transverse
momentum of the pion pair $\bk_T$), transverse ``side" (in the transverse plane
and perpendicular to $\bk_T$), and longitudinal (``long") directions,
respectively.  In Eq.~(\ref{Eq-CF}), $\lambda$ is the chaoticity parameter,
and $R_{\rm out}$, $R_{\rm side}$, and $R_{\rm long}$ are the HBT radii in
the out, side, and long directions respectively.

In experimental HBT analyses, the Coulomb effect on the correlation functions
must be considered \cite{Gyu79,Pratt86,YMLiu86,Lisa05}.  It becomes the
well-known Gamow factor \cite{Schiff} in the case of first-order approximation.
Because the Coulomb effect has been carefully removed  from the experimental HBT
data \cite{STAR-HBT-prc05,ALI-HBT-prc16}, we will not consider it in our model
HBT analyses.  In a more general case, the interference term in the formula of
the two-pion correlation function also contains factors concerning two relative
momentum products, $e^{-2R_{\rm os}^2 q_{\rm out} q_{\rm side}}$, $e^{-2R_{\rm ol}^2
q_{\rm out} q_{\rm long}}$, and $e^{-2R_{\rm sl}^2 q_{\rm side} q_{\rm long}}$ \cite{ChapScotHeinz95,ChapNixHeinz95,E895-prl00,E895-plb00,Wie99,Lisa05,STAR-HBT-prc05}.
They are employed in more detailed HBT analyses.
In Ref. \cite{STAR-HBT-prc05}, the STAR collaboration investigates the dependence
of two-pion correlation functions on the particle azimuth angle using a formula
containing the factor $e^{-2R_{\rm os}^2 q_{\rm out} q_{\rm side}}$, and the relevant
Fourier coefficients of HBT correlations are studied as a function of the number of
participating nucleons in heavy-ion collisions at RHIC energy.
In Ref. \cite{ALI-HBT-prc16}, the ALICE collaboration performs two-pion HBT
analyses with the correlation-function formula without the factors concerning
two relative momentum products in heavy-ion collisions at LHC energy.
Because the experimental data of HBT radii $R_{\rm out}$, $R_{\rm side}$, and
$R_{\rm long}$ are available in heavy-ion collisions at both the RHIC and
LHC, we perform the HBT analyses with formula (\ref{Eq-CF}) for the granular
sources at RHIC and LHC energies and examine the variations of model
parameters with collision energy.  Further detailed HBT analyses with the more
general correlation-function formula for the granular sources will be of interest.

In Fig. \ref{Figgs-hbt} we show the results of pion interferometry of the
granular sources for Au-Au collisions  at $\sqrt{s_{NN}}=200$ GeV and
Pb-Pb at $\sqrt{s_{NN}}=2.76$ TeV in the corresponding centrality intervals
as in Fig. \ref{Figgs-spe}.  The experimental data of the pion
interferometry analyses performed by the STAR \cite{STAR-HBT-prc05} and
the ALICE \cite{ALI-HBT-prc16} collaborations are also shown in Fig.
\ref{Figgs-hbt} by the black solid circles and black solid squares,
respectively.  In our interferometry analyses for the granular sources, we
take the same rapidity or pseudorapidity cuts as in the experimental analyses
\cite{STAR-HBT-prc05,ALI-HBT-prc16}.  In Fig. \ref{Figgs-hbt}, the results
denoted by cyan open circles are for the granular sources without viscosity
and with the same model parameters as in Ref. \cite{YangZhang15-Gyongyos}
for the Woods-Saxon initial energy distribution of the droplet.  The results
denoted by red down-triangles and blue up-triangles are for the granular
sources with only shear viscosity and with both shear and bulk viscosities,
respectively.  The parameters for the viscous sources are taken to be the same
as for the non-viscous sources \cite{YangZhang15-Gyongyos}.  One can see from
Fig. \ref{Figgs-hbt} that the differences between the results for the granular
sources with only shear viscosity and with both shear and bulk viscosities are
very small.  The main influence of viscosity is on the HBT radius $R_{\rm out}$,
which has an obvious decrease for the viscous sources.  The effect of viscosity
leads also to the decrease of the ratio $R_{\rm out}/R_{\rm side}$.  The results
of the chaotic parameters $\lambda$ of the granular sources are larger than the
experimental data, because many effects in experiments can decrease the measurement
value of $\lambda$ \cite{Gyu79,Wongbook,Wie99,Wei00,Lisa05}, which exceed our
considerations in the granular source model.

\end{multicols}
\begin{center}
\includegraphics[width=15cm]{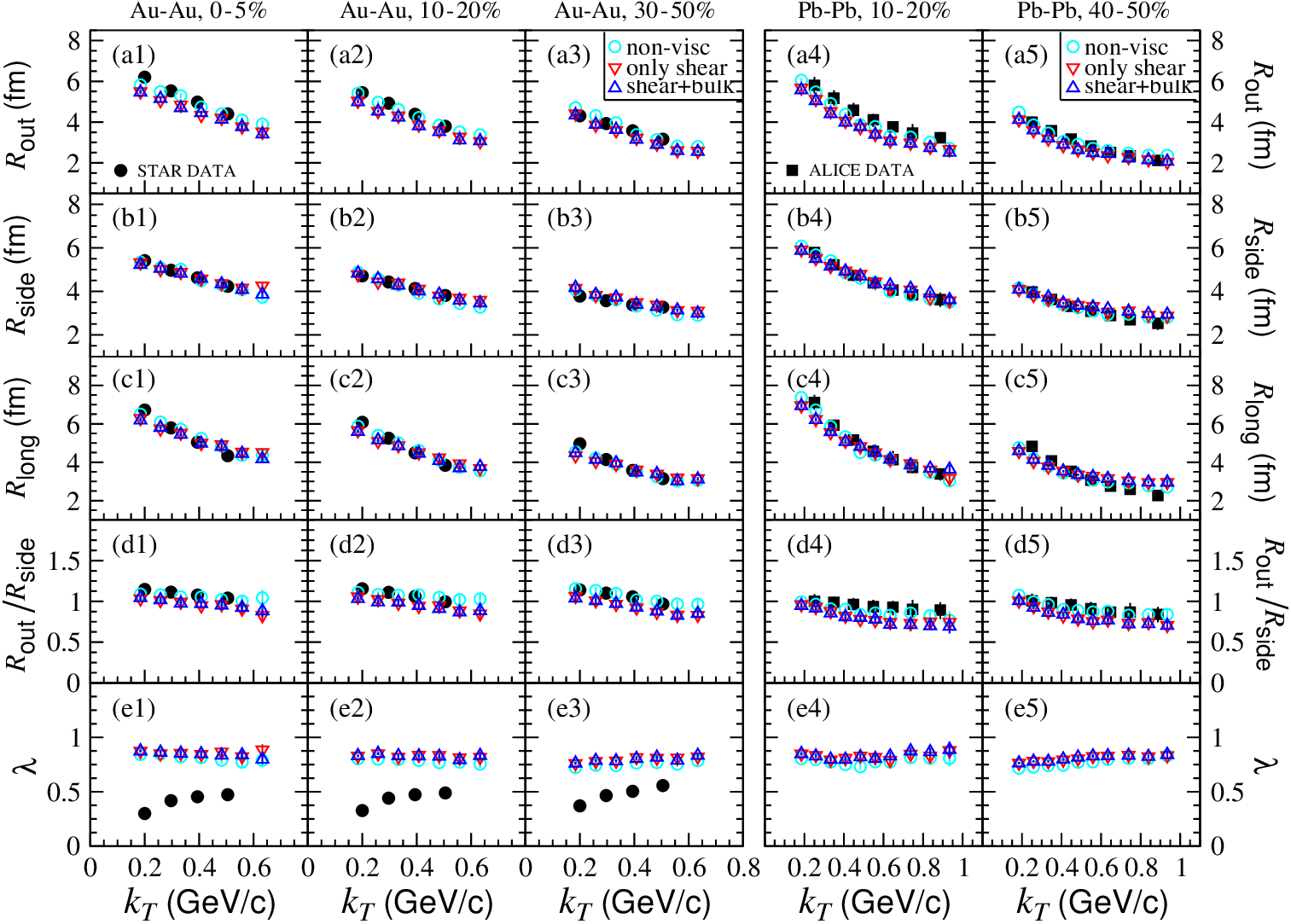}
\vspace*{3mm}
\figcaption{\label{Figgs-hbt} (Color online)
Pion HBT radii and chaotic parameter of the granular
sources without viscosity (cyan open circles), with only shear viscosity (red
down-triangles), and with both shear and bulk viscosities (blue up-triangles),
for Au-Au collisions at $\sqrt{s_{NN}}=200$ GeV and Pb-Pb collisions
at $\sqrt{s_{NN}}=2.76$ TeV in the same centrality intervals as in Fig.
\ref{Figgs-spe}.  The black solid circles in the left three
columns show the experimental data for Au-Au collisions measured by
the STAR collaboration \cite{STAR-HBT-prc05}, and the black solid
squares in the right two columns show the experimental data for Pb-Pb
collisions measured by the ALICE collaboration \cite{ALI-HBT-prc16}.}
\end{center}
\begin{multicols}{2}

\section{Influence of viscosity on granular source parameters}
We saw in the last section that the pion transverse-momentum spectrum and elliptic
flow change little for the granular source with and without viscosity.  However,
the pion HBT radius $R_{\rm out}$ of the viscous granular source is obviously
smaller than that of the granular source without viscosity.  Because $R_{\rm out}$
is related not only to the source space-geometry but also to the source evolution
time and expansion velocity in high-energy heavy-ion collisions \cite{Wie99,HerBer95,ChapScotHeinz95,YinYangZhang-prc12,YangZhang14-ahep}, it is
sensitive to the geometry as well as velocity parameters in the granular source
model.  The model parameters we used in the last section are fixed for the granular
sources without viscosity \cite{YangZhang15-Gyongyos}.  In this section, we will
determine the model parameters for the viscous granular sources by the
experimental data of pion HBT radii as well as pion transverse-momentum
spectra and elliptic flow.

\vspace{0mm}
\end{multicols}
\begin{center}
\tabcaption{ \label{Tab-parameter}  Adjusted model parameter values and their
original values in Ref. \cite{YangZhang15-Gyongyos}.}
\footnotesize
\begin{tabular*}{170mm}{@{\extracolsep{\fill}}c|cccc|cccc}
\toprule
 & \multicolumn{4}{c|}{Set II} & \multicolumn{4}{c}{Set I \cite{YangZhang15-Gyongyos}}\\
 &~$\sigma_d\,\text{(fm)}~$~&~$\Delta {\cal R}_T\,
\text{(fm)}~$&~~~~$\overline{a}_T$~~~&~~~~~$\Delta a_T$~~&
~$\sigma_d\,\text{(fm)}~$~&~$\Delta {\cal R}_T\,\text{(fm)}~$&
~~~~$\overline{a}_T$~~~&~~~~~$\Delta a_T$~~\\
\hline
~Au-Au,~~0--5~\%&3.0&0.21&~0.445&~~~0.069&2.5&0.70&~0.469&~~~0.066 \\
~Au-Au,~10--20\%&2.8&0.15&~0.434&~~~0.126&2.5&0.50&~0.457&~~~0.122 \\
~Au-Au,~30--50\%&2.5&0.00&~0.408&~~~0.167&2.5&0.30&~0.453&~~~0.156 \\
~Pb-Pb,~10--20\%&3.0&0.27&~0.471&~~~0.095&2.5&0.90&~0.496&~~~0.092 \\
~Pb-Pb,~40--50\%&2.5&0.00&~0.412&~~~0.130&2.5&0.40&~0.434&~~~0.127 \\
\bottomrule
\end{tabular*}%
\end{center}
\begin{multicols}{2}

As viscosity speeds up droplet evolution, on average, we increase the
initial droplet radius parameter $\sigma_d$ and decrease the initial droplet
velocity parameter $\overline{a}_T$ for the viscous granular source compared
to those for the granular source without viscosity, to counteract the effects
of viscosity on the source evolution.  The increase of $\sigma_d$ leads to
an increase of the HBT radius $R_{\rm out}$ and also an increase of the HBT
radius $R_{\rm side}$.
We further decrease the shell parameter $\Delta {\cal R}_T$ for the viscous
granular source to decrease the value of $R_{\rm side}$, and adjust the velocity
parameters $\overline{a}_T$ and $\Delta a_T$ somewhat to let the results of the
momentum spectrum and elliptic flow of the viscous granular sources conform to
the experimental data.  The other model parameters used for the
viscous granular sources are the same as in Ref. \cite{YangZhang15-Gyongyos}
for the granular sources without viscosity.  In Table \ref{Tab-parameter}, we
present the adjusted model parameter values (Set II) for the viscous granular
sources for heavy-ion collisions of Au-Au at $\sqrt{s_{NN}} =200$ GeV in
the centrality intervals 0-5\%, 10-20\%, 30-50\%, and for heavy-ion
collisions of Pb-Pb at $\sqrt{s_{NN}}=2.76$ TeV in the centrality intervals
10-20\% and 40-50\%.  Meanwhile, the corresponding model parameter values in
Ref. \cite{YangZhang15-Gyongyos} (Set I), determined for the granular sources
without viscosity, are also presented for comparison.

\end{multicols}
\vspace*{1mm}
\begin{center}
\includegraphics[width=15cm]{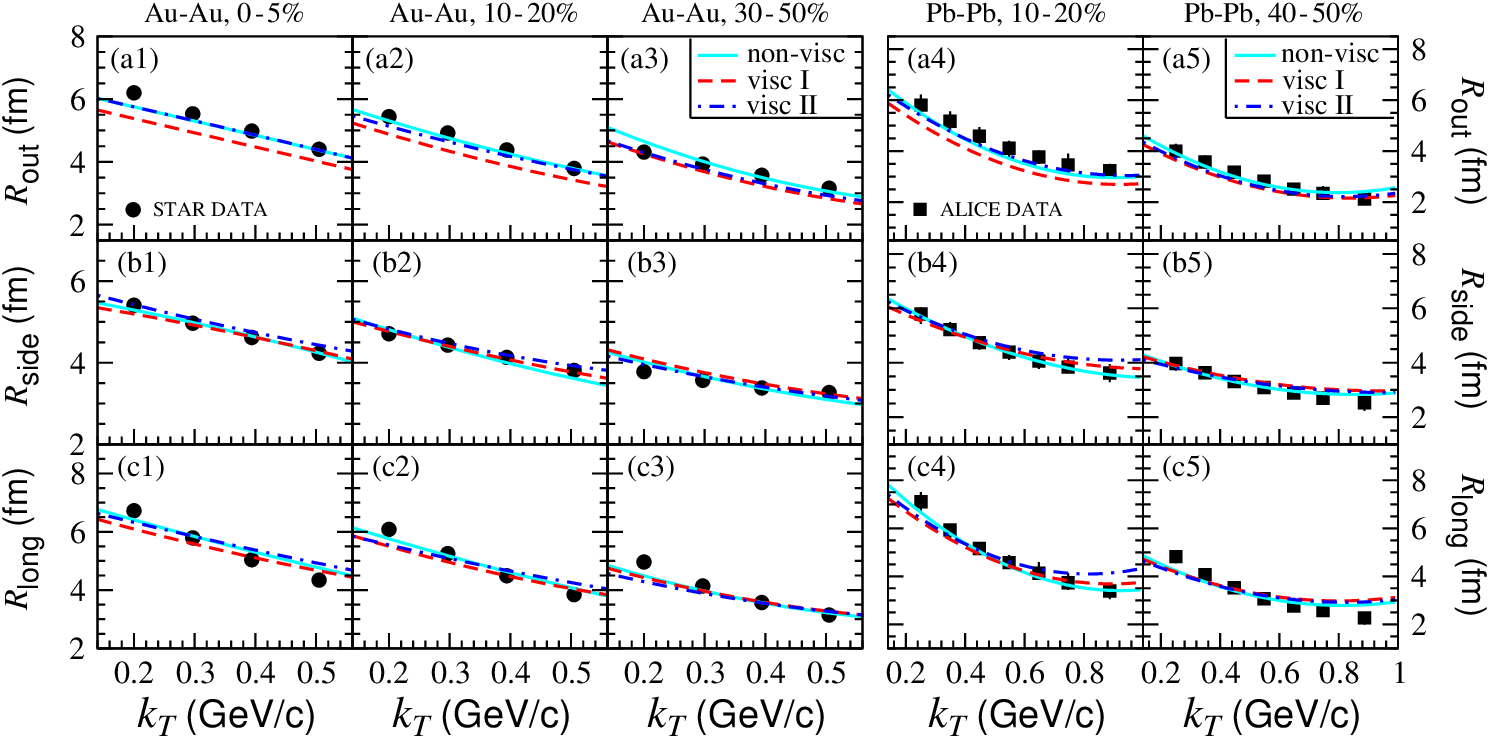}
\vspace*{3mm}
\figcaption{\label{Figgs-hbt-line} (Color online)
Pion HBT radii of granular sources without
viscosity (cyan solid lines) and with viscosity (red dashed and blue
dot-dashed lines), for Au-Au collisions at $\sqrt{s_{NN}}=200$ GeV and
Pb-Pb collisions at $\sqrt{s_{NN}}=2.76$ TeV in the same centrality
intervals as in Figs. \ref{Figgs-spe} -- \ref{Figgs-hbt}.  The model
parameters used for the solid and dashed lines are the same as in Ref.
\cite{YangZhang15-Gyongyos} (parameter set I), and the dot-dashed lines
are for the results of viscous granular sources with the adjusted model
parameters (parameter set II).  The experimental results of the HBT radii of
Au-Au collisions measured by the STAR collaboration \cite{STAR-HBT-prc05},
and Pb-Pb collisions measured by the ALICE collaboration
\cite{ALI-HBT-prc16} are also plotted.}
\end{center}
\begin{multicols}{2}

In Fig.~\ref{Figgs-hbt-line}, we compare the pion HBT radii of viscous
granular sources with the adjusted model parameters (visc-II) and with the
model parameters as in Ref. \cite{YangZhang15-Gyongyos} for granular
sources without viscosity (visc-I). We also compare with the results of
the granular source without viscosity (non-visc) and the experimental data
of Au-Au collisions at $\sqrt{s_{NN}}=200$ GeV \cite{STAR-HBT-prc05}
and Pb-Pb collisions at $\sqrt{s_{NN}}=2.76$ TeV \cite{ALI-HBT-prc16}
in the same centrality intervals as in Figs. \ref{Figgs-spe} --
\ref{Figgs-hbt}.  Here, the lines are obtained by fitting the separated
results of HBT radii $R(k_T)$ of the granular sources with a binomial
of $k_T$.  The HBT results of $R_{\rm out}$ for the
viscous granular sources with the adjusted parameters are larger than
those with the original model parameters \cite{YangZhang15-Gyongyos}
and closer to the experimental results.  There is not much difference in the results of the HBT
radii $R_{\rm side}$ and $R_{\rm long}$ for the two sets of model
parameters.

We further examine the pion transverse-momentum spectrum and elliptic flow
of the granular sources with the model parameter sets I and II.  In Fig.
\ref{Figgs-spe-v2n}, we show the results of pion transverse-momentum spectrum
[in panel (a)] and elliptic flow [in panels (b) and (c)] of the granular
sources with and without viscosity, for Au-Au collisions at $\sqrt{s_{NN}}
=200$ GeV in the centrality intervals 0-5\%, 10-20\%, and 30-50\%, and Pb-Pb collisions at $\sqrt{s_{NN}}=2.76$ TeV in the centrality intervals 10-20\%
and 40-50\%.  Here, the cyan solid lines are for granular sources without
viscosity and calculated with model parameter set I, the same as in Ref.
\cite{YangZhang15-Gyongyos}.  The red dashed and blue dot-dashed lines are
for viscous granular sources with model parameter sets I and II,
respectively.  The experimental data of the momentum spectrum and elliptic
flow measured by the PHENIX and STAR collaborations at the RHIC
\cite{PHE-spe04z,STA-spe04z,STA-v2-05z}, and by the ALICE collaboration at the
LHC \cite{ALI-v2-11z,ALI-v2-13z} are also plotted.  At lower $p_T$, the
results of the momentum spectra (or elliptic flow) of the granular sources
with and without viscosity have little difference, and they are agree with
the experimental data.  The differences between the granular source results
and the experimental data at larger $p_t$ are due to the absence of hard
processes in the granular source models.

\end{multicols}
\vspace*{1mm}
\begin{center}
\includegraphics[scale=0.5]{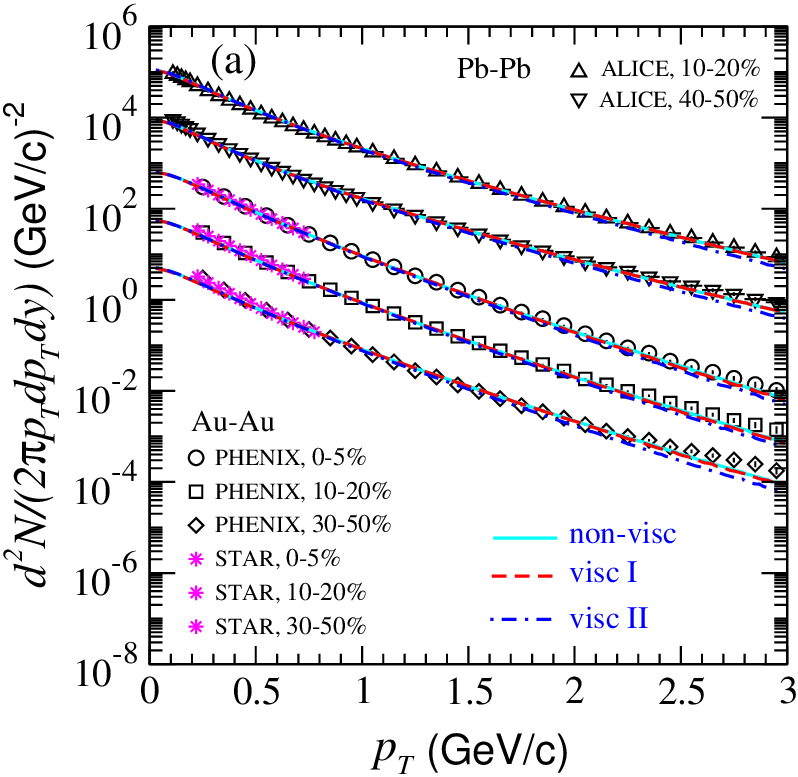}
\vspace*{3mm}
\includegraphics[scale=0.5]{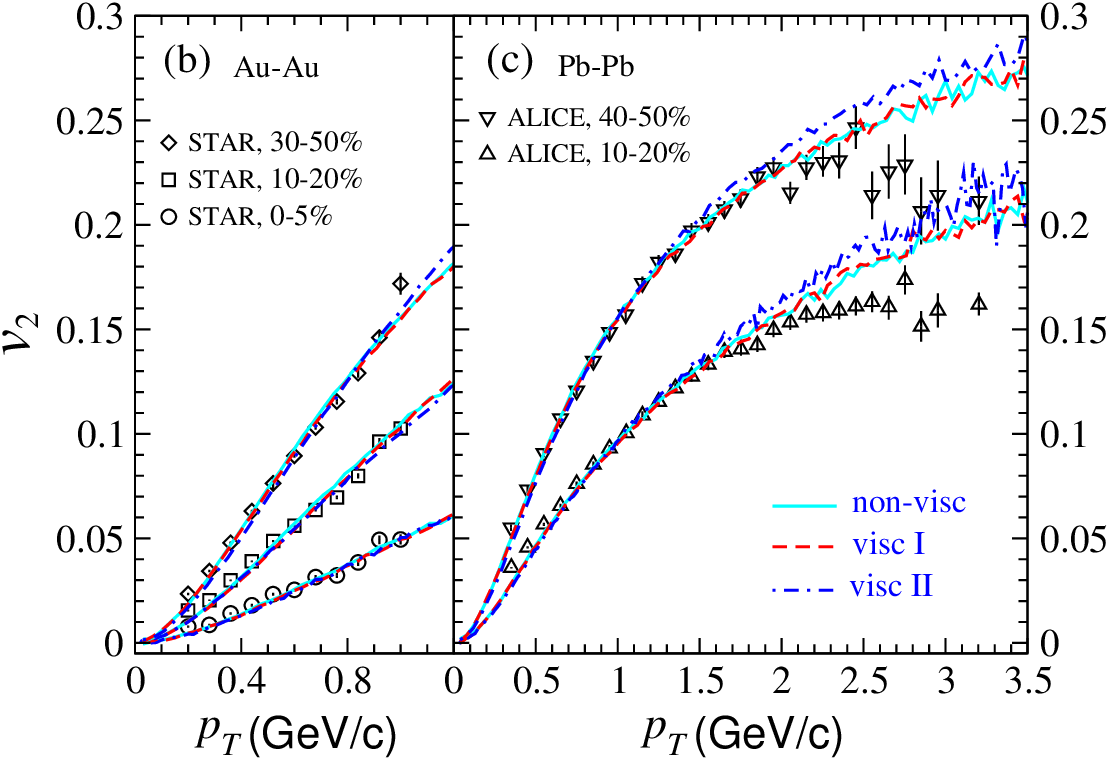}
\figcaption{\label{Figgs-spe-v2n} (Color online)
(a) Pion transverse momentum spectra of granular sources without
viscosity (cyan solid lines) and with viscosity (red dashed lines and
blue dot-dashed lines), for Au-Au collisions at $\sqrt{s_{NN}}=200$
GeV and Pb-Pb collisions at $\sqrt{s_{NN}}=2.76$ TeV.  The model
parameters used for the granular sources without viscosity are the same
as in Ref. \cite{YangZhang15-Gyongyos} (parameter set I), and the
parameters used for the viscous granular sources are the same as in Ref.
\cite{YangZhang15-Gyongyos} (parameter set I) and the adjusted model
parameters (parameter set II), respectively.  The experimental data
measured by the PHENIX and the STAR collaborations at the RHIC
\cite{PHE-spe04z,STA-spe04z} and by the ALICE collaboration
at the LHC \cite{ALI-spe13z} are also plotted.
(b) and (c) Pion elliptic flow of the granular sources without viscosity
(cyan solid lines) and with viscosity (red dashed lines for model parameter
set I and blue dot-dashed lines for model parameter set II)), for Au-Au
collisions at $\sqrt{s_{NN}}=200$ GeV and Pb-Pb collisions at
$\sqrt{s_{NN}}=2.76$ TeV, respectively.  The experimental data measured
by the STAR collaboration at the RHIC \cite{STA-v2-05z} and by
the ALICE collaboration at the LHC \cite{ALI-v2-11z,ALI-v2-13z} are also
plotted in panels (b) and (c), respectively.}
\end{center}
\begin{multicols}{2}

In early articles on viscous hydrodynamics, it was pointed out that shear
viscosity may lead to the increase in the particle transverse-momentum spectrum
at high $p_T$ \cite{MurRis04,BaiRomWie06,DusTea-prc08,SonHei-prc10}, and to the
decrease of elliptic flow in high $p_T$ regions compared to the ideal hydrodynamic
results \cite{Rom-PRL07,DusTea-prc08,SonHei08,LuzRom-prc08,SonHei-prc10}.
Also, it was expected that the viscosity might increase the HBT radius
$R_{\rm out}$ and decrease slightly the HBT radius $R_{\rm side}$
\cite{MurRis04,PraVre08}.  In the last few years, many groups have utilizes the
viscosity effects on the momentum spectrum and elliptic flow to investigate
the QGP viscosities, by comparing their model results with experimental
data from relativistic heavy-ion collisions \cite{{s95p-Shen10,SchJeoGal-PRL11,
Song-prc11,HeiSne13,Rugg14,Song-prc11a,Shen-prc11,SchenkeJeonGale-plb11,
ShenHeinz-prc12,Soltz13,Karpenko15}}.
In Ref. \cite{Bozek11}, P. Bo\.{z}ek investigated the HBT radii for central heavy-ion collisions in a viscous hydrodynamic model and found that
the initial pre-equilibrium flow reduces the HBT radius $R_{\rm out}$.  It
has also been found that the bulk viscosity has very small influence on HBT radii
\cite{BozekWysk12}.

In the viscous granular source model considered, the outward motion of the droplet
reduces the effect of viscosity on transverse-momentum spectra, as mentioned in
Sec. 3.  Because the elliptic flow in the granular source model is determined by
the anisotropic initial velocity of the droplets, the influences of viscosity
on the $v_2$ values are negligible.  As a result, the differences in the model
velocity parameters between granular sources with and without viscosity
are very small.  However, viscosity may increase the HBT radius $R_{\rm out}$
in the granular source model and therefore lead to changes in the geometry
parameters of the viscous granular sources from those of the granular sources
without viscosity.
One can see from Table \ref{Tab-parameter} that the values of $\sigma_d$ for
the viscous granular sources are larger than those for the granular sources
without viscosity, except for the most peripheral collisions.  Also, the values
of the shell parameter of source $\Delta {\cal R}_T$ of the viscous granular
sources are smaller than those of the granular sources without viscosity.
These indicate that the viscous granular sources have larger droplets and a less
obvious shell-like structure than the granular sources without viscosity.
The average number of droplets in a granular source is related to the initial
source volume ${\cal R}_T^2 {\cal R}_Z$, the shell parameter $\Delta {\cal R}_T$,
and $\sigma_d$.  For a uniform distribution of the droplets in a granular source,
the average number of droplets in the source is proportional to $({\cal R}_T^2
{\cal R}_z/\langle a^3\rangle)$ $\propto ({\cal R}_T^2{\cal R}_z/\sigma_d^3)$,
where $\langle a^3\rangle$ denotes the average of the droplet-radius cube over
granular sources with the same collision energy and centrality.

\section{Summary and Discussion}
Based on the second-order M\"{u}ller-Israel-Stewart theory of dissipative
hydrodynamics \cite{2ndMIS}, we have given a viscous hydrodynamic description
for a spherical QGP droplet and examined its evolution.  A granular source
model consisting of the viscous QGP droplets has been developed.  We have analyzed pion
transverse-momentum spectra, elliptic flow, and Hanbury-Brown-Twiss (HBT)
interferometry in the viscous granular source model, for collisions of
Au-Au at $\sqrt{s_{NN}}=200$ GeV and Pb-Pb at $\sqrt{s_{NN}}=2.76$ TeV and
in different centrality intervals.  By comparing the multi-observable
results of the granular sources with the experimental data, we have investigated
the effects of viscosity on the observables.  We find that the shear viscosity
of the QGP droplet speeds up the droplet evolution and the influence of the
bulk viscosity on the evolution is negligible.  Although there are viscous
effects on the droplet evolution, the pion transverse-momentum spectrum and
elliptic flow change little for granular sources with and without
viscosity.  This is because the outward motion of droplets in the granular
sources reduce the effect of viscosity on the pion momentum spectrum, and
the value of elliptic flow is mainly determined by the anisotropic initial
velocities of droplets in the granular source model.  The crucial effect
of droplet viscosity is on the HBT radius $R_{\rm out}$.  It has a significant
decrease for viscous granular sources compared to granular
sources without viscosity.

In high-energy heavy-ion collisions, transverse-momentum spectra, elliptic flow,
and HBT correlations are important observables.  The experimental data of these
observables give strict constraints on the model parameters.  Examining the changes in model
parameters  with collision energy, centrality, and viscosity may provide
an integrated picture of source geometry and dynamics.  We determine the model
parameters of the viscous granular sources by the experimental data of these
observables for pions and investigate the effects of viscosity on the
model parameters.  The results indicate that the viscosity of the droplet leads to
an increase in the initial droplet-radius parameter and decreases in the
source-shell parameter and the initial droplet velocity in the granular
source model.  These change the initial picture of a granular source
relative to the case without viscosity.

As a simplification of the particle-emitting sources with granular inhomogeneous
structures formed in ultrarelativistic heavy-ion collisions, granular source
models are used to describe the system evolution after a later stage of QGP
expansion.  In the models, the QGP lumps are assumed to have spherical geometry
for simplicity, and anisotropic pressure in the early QGP matter is assumed
to lead to anisotropic initial velocities of the droplets.
We determine the model geometry and velocity parameters $({\cal R}_T, {\cal R}_z,
\sigma_d)$ and $(\overline{a}_T, \Delta a_T, a_z, b_T, b_z)$ by the experimental
data of the multi-observables in order to let the model reproduce these experimental
observables.  The variations of the model parameters with collision
centrality and energy exhibit certain regularities.
The results of the
pion transverse-momentum spectra, elliptic flow, and HBT radii of the granular
sources are in accordance with the experimental data of Au-Au collisions at
$\sqrt{s_{NN}} =200$ GeV and Pb-Pb collisions at $\sqrt{s_{NN}}=2.76$ TeV
in different centrality intervals.  This means that the granular source model,
with several evolving centers as the main characteristic, reflects in some
degree the physics of the system evolution during the later stages of
heavy-ion collisions.  The model parameters of the granular sources determined
by the experimental data are helpful for the study of the QGP properties and
dynamics at earlier stages.  Finally, it should be mentioned that the present
granular source models involve neither the possible overlap of the droplets
evolving later, nor particle rescattering and absorption by other
droplets in the source.  Further improvements of the granular source model
to make it more realistic, and applying it to other hadron observable analyses, such as kaons, will be of interest.

\acknowledgments{The authors would like to thank M. J. Efaaf, Z. Q. Su, L. Cheng,
and H. C. Song for their helpful discussions.}

\end{multicols}

\vspace{15mm}

\begin{multicols}{2}

\end{multicols}
\subsection*{Appendix A: Derivation for relaxation equation}

\begin{small}
\begin{subequations}
\renewcommand{\theequation}{A\arabic{equation}}

Here, we give details of how to derive the relaxation equations for shear
viscosity in Eq.\,(\ref{single-evol}) from Eq.\,(\ref{2ndMIS-1}). \\

\noindent{\bf 1. First term of Eq.\,(\ref{2ndMIS-1})}\\

The first term of Eq.\,(\ref{2ndMIS-1}) can be expanded as
\begin{equation}
\begin{aligned}
\tau_{\pi}\Delta^{\mu\alpha}\Delta^{\nu\beta}D\pi_{\alpha\beta}
&=\tau_{\pi}(g^{\mu\alpha}-u^\mu u^\alpha)(g^{\nu\beta}-u^\nu u^\beta)D\pi_{\alpha\beta}\\
&=\tau_{\pi}g^{\mu\alpha}g^{\nu\beta}D\pi_{\alpha\beta}
  -\tau_{\pi}g^{\mu\alpha}u^{\nu}u^{\beta}D\pi_{\alpha\beta}
  -\tau_{\pi}u^{\mu}u^{\alpha}g^{\nu\beta}D\pi_{\alpha\beta}
  -\tau_{\pi}u^{\mu}u^{\alpha}u^{\nu}u^{\beta}D\pi_{\alpha\beta}.
\end{aligned}
\end{equation}
With the relations $Dg_{\mu\nu}=Dg^{\mu\nu}=0$ and the orthogonality $u_\mu\pi^{\mu\nu}=u^\mu\pi_{\mu\nu}=0$, we have
\begin{equation}
\left\{
\begin{aligned}
g^{\mu\alpha}g^{\nu\beta}D\pi_{\alpha\beta}
&=g^{\nu\beta}[D(g^{\mu\alpha}\pi_{\alpha\beta})-(Dg^{\mu\alpha})\pi_{\alpha\beta}]
 =g^{\nu\beta}(D\pi_{~\beta}^{\mu}),\\
g^{\mu\alpha}u^{\nu}u^{\beta}D\pi_{\alpha\beta}
&=g^{\mu\alpha}u^{\nu}[D(u^{\beta}\pi_{\alpha\beta})-(Du^{\beta})\pi_{\alpha\beta}]
 =-g^{\mu\alpha}u^{\nu}(Du^\beta)\pi_{\alpha\beta},\\
u^{\mu}u^{\alpha}g^{\nu\beta}D\pi_{\alpha\beta}
&=g^{\nu\beta}u^{\mu}[D(u^{\alpha}\pi_{\alpha\beta})-(Du^{\alpha})\pi_{\alpha\beta}]
 =-g^{\nu\beta}u^{\mu}(Du^{\alpha})\pi_{\alpha\beta},\\
u^{\mu}u^{\alpha}u^{\nu}u^{\beta}D\pi_{\alpha\beta}
&=u^{\mu}u^{\nu}[u^{\alpha}D(u^{\beta}\pi_{\alpha\beta})-(Du^{\beta})u^{\alpha}\pi_{\alpha\beta}]
 =0.
\end{aligned}
\right.
\end{equation}
Therefore, the first term of Eq.\,(\ref{2ndMIS-1}) can be expressed as
\begin{equation}
\begin{aligned}
\tau_{\pi}\Delta^{\mu\alpha}\Delta^{\nu\beta}D\pi_{\alpha\beta}=\tau_{\pi}[g^{\nu\beta}(D\pi_{~\beta}^{\mu})+I^{\mu\nu}],
\end{aligned}
\end{equation}
where
\begin{equation}
I^{\mu\nu}\,{\equiv}\,g^{\mu\alpha}u^{\nu}(Du^\beta)\pi_{\alpha\beta}+g^{\nu\beta}u^{\mu}(Du^{\alpha})\pi_{\alpha\beta}.
\end{equation}

For a symmetrical sphere, $u^\theta=u^\phi=0$, so $I^{\theta\theta}=I^{\phi\phi}=0$.
The non-vanishing components of $I^{\mu\nu}$ are as follows,
\begin{equation}
\left\{
\begin{aligned}
I^{tt}
&=g^{t\alpha}u^{t}(Du^\beta)\pi_{\alpha\beta}+g^{t\beta}u^{t}(Du^{\alpha})\pi_{\alpha\beta}\\
&=2g^{t\alpha}u^{t}(Du^\beta)\pi_{\alpha\beta}
 =2g^{tt}u^{t}(Du^t)\pi_{tt}+2g^{tt}u^{t}(Du^r)\pi_{tr}
 =-2\gamma^4v_r(Dv_r)\tau^{rr},\\
I^{tr}
&=g^{t\alpha}u^{r}(Du^\beta)\pi_{\alpha\beta}+g^{r\beta}u^{t}(Du^{\alpha})\pi_{\alpha\beta}\\
&=g^{tt}u^{r}(Du^t)\pi_{tt}+g^{tt}u^{r}(Du^r)\pi_{tr}+g^{rr}u^{t}(Du^t)\pi_{tr}+g^{rr}u^{t}(Du^r)\pi_{rr}
 =-\gamma^4[v_r^2+1](Dv_r)\tau^{rr},\\
I^{rt}
&=g^{r\alpha}u^{t}(Du^\beta)\pi_{\alpha\beta}+g^{t\beta}u^{r}(Du^{\alpha})\pi_{\alpha\beta}
 =I^{tr},\\
I^{rr}
&=g^{r\alpha}u^{r}(Du^\beta)\pi_{\alpha\beta}+g^{r\beta}u^{r}(Du^{\alpha})\pi_{\alpha\beta}\\
&=2g^{r\alpha}u^{r}(Du^\beta)\pi_{\alpha\beta}
 =2g^{rr}u^{r}(Du^t)\pi_{rt}+2g^{rr}u^{r}(Du^r)\pi_{rr}
 =-2\gamma^4v_r(Dv_r)\tau^{rr}=I^{tt}.
\end{aligned}
\right.
\end{equation}
Here we use the relations
\begin{equation}
\gamma=\frac{1}{\sqrt{1-v_r^2}}
,~~
v_r\frac{\partial(\gamma v_r)}{\partial t}=\frac{\partial \gamma}{\partial t}
,~~
v_r\frac{\partial(\gamma v_r)}{\partial r}=\frac{\partial \gamma}{\partial r}.
\end{equation}
Therefore, we have
\begin{equation}
\begin{aligned}
\tau_{\pi}\Delta^{t\alpha}\Delta^{t\beta}D\pi_{\alpha\beta}
&=\tau_{\pi}[g^{t\beta}(D\pi_{~\beta}^{t})+I^{tt}]
 =\tau_{\pi}[g^{tt}(D\pi_{~ t}^{t})+I^{tt}]
 =\tau_{\pi}\gamma^2v_r^2(D\tau^{rr}),\\
\tau_{\pi}\Delta^{t\alpha}\Delta^{r\beta}D\pi_{\alpha\beta}
&=\tau_{\pi}[g^{t\beta}(D\pi_{~\beta}^{r})+I^{tr}]
 =\tau_{\pi}[g^{tt}(D\pi_{~ t}^{r})+I^{tr}]
 =\tau_{\pi}\gamma^2v_r(D\tau^{rr}),\\
\tau_{\pi}\Delta^{r\alpha}\Delta^{t\beta}D\pi_{\alpha\beta}
&=\tau_{\pi}[g^{r\beta}(D\pi_{~\beta}^{t})+I^{rt}]
 =\tau_{\pi}[g^{rr}(D\pi_{~ r}^{t})+I^{rt}]
 =\tau_{\pi}\Delta^{t\alpha}\Delta^{r\beta}D\pi_{\alpha\beta},\\
\tau_{\pi}\Delta^{r\alpha}\Delta^{r\beta}D\pi_{\alpha\beta}
&=\tau_{\pi}[g^{r\beta}(D\pi_{~\beta}^{r})+I^{rr}]
 =\tau_{\pi}[g^{rr}(D\pi_{~ r}^{r})+I^{rr}]
 =\tau_{\pi}\gamma^2(D\tau^{rr}),\\
\tau_{\pi}\Delta^{\theta\alpha}\Delta^{\theta\beta}D\pi_{\alpha\beta}
&=\tau_{\pi}[g^{\theta\beta}(D\pi_{~\beta}^{\theta})]
 =\tau_{\pi}[g^{\theta\theta}(D\pi_{~\theta}^{\theta})]
 =\tau_{\pi}\frac{1}{r^2}(D\tau^{\theta\theta}),\\
\tau_{\pi}\Delta^{\phi\alpha}\Delta^{\phi\beta}D\pi_{\alpha\beta}
&=\tau_{\pi}[g^{\phi\beta}(D\pi_{~\beta}^{\phi})]
 =\tau_{\pi}[g^{\phi\phi}(D\pi_{~\phi}^{\phi})]
 =\tau_{\pi}\frac{1}{r^2\sin^2\theta}(D\tau^{\phi\phi}),
\end{aligned}
\end{equation}
where $D=\gamma(\partial_t+v_r\partial_r)$.\\

\noindent{\bf 2. Second term of Eq.\,(\ref{2ndMIS-1})}\\

For a spherical system, we have

\begin{equation}
\begin{aligned}
&\pi^{\mu\nu}
=\left(
   \begin{array}{cccc}
   \gamma^2v_r^2\tau^{tt} & \gamma^2v_r\tau^{tr} & 0                                & 0                                        \\
   \gamma^2v_r\tau^{rt}   & \gamma^2\tau^{rr}    & 0                                & 0                                        \\
   0                      & 0                    & \frac{1}{r^2}\tau^{\theta\theta} & 0                                        \\
   0                      & 0                    & 0                                & \frac{1}{r^2\sin^2\theta}\tau^{\phi\phi} \\
   \end{array}
 \right),
\end{aligned}
\end{equation}
where $\tau^{tt}=\tau^{tr}=\tau^{rt}=\tau^{rr}$ and $\tau^{\theta\theta}=\tau^{\phi\phi}$.
\begin{equation}
\begin{aligned}
\pi_{\alpha\beta}
=&g_{\alpha\mu}\pi^{\mu\nu}g_{\nu\beta}\\
=&\left(
   \begin{array}{cccc}
   1 &  0 &  0   &  0               \\
     & -1 &  0   &  0               \\
     &  0 & -r^2 &  0               \\
     &  0 &  0   & -r^2\sin^2\theta \\
   \end{array}
  \right)
  \left(
   \begin{array}{cccc}
   \gamma^2v_r^2\tau^{rr} & \gamma^2v_r\tau^{rr} & 0                                & 0 \\
   \gamma^2v_r\tau^{rr}   & \gamma^2v_r\tau^{rr} & 0                                & 0 \\
   0                      & 0                    & \frac{1}{r^2}\tau^{\theta\theta} & 0 \\
   0                      & 0                    & 0                                & 0 \frac{1}{r^2\sin^2\theta}\tau^{\phi\phi} \\
   \end{array}
 \right)
 \left(
   \begin{array}{cccc}
   1 &  0 &  0   &  0               \\
     & -1 &  0   &  0               \\
     &  0 & -r^2 &  0               \\
     &  0 &  0   & -r^2\sin^2\theta \\
   \end{array}
 \right)\\
=&\left(
   \begin{array}{cccc}
   \gamma^2v_r^2\tau^{rr}  & -\gamma^2v_r\tau^{rr} & 0                      & 0                              \\
   -\gamma^2v_r\tau^{rr}   & \gamma^2\tau^{rr}     & 0                      & 0                              \\
   0                       & 0                     & r^2\tau^{\theta\theta} & 0                              \\
   0                       & 0                     & 0                      & r^2\sin^2\theta\tau^{\phi\phi} \\
   \end{array}
 \right),
\end{aligned}
\end{equation}
\begin{equation}
\begin{aligned}
\pi_{~\nu}^{\mu}
&=g^{\mu\lambda}\pi_{\lambda\nu}\\
&=\left(
    \begin{array}{cccc}
     1 &  0 &  0             &  0                         \\
       & -1 &  0             &  0                         \\
       &  0 & -\frac{1}{r^2} &  0                         \\
       &  0 &  0             & -\frac{1}{r^2\sin^2\theta} \\
    \end{array}
  \right)
  \left(
    \begin{array}{cccc}
       \gamma^2v_r^2\tau^{rr} & -\gamma^2v_r\tau^{rr} & 0                      & 0                              \\
      -\gamma^2v_r\tau^{rr}   &  \gamma^2\tau^{rr}    & 0                      & 0                              \\
       0                      &  0                    & r^2\tau^{\theta\theta} & 0                              \\
       0                      &  0                    & 0                      & r^2\sin^2\theta\tau^{\phi\phi} \\
    \end{array}
  \right)\\
&=\left(
    \begin{array}{cccc}
      \gamma^2v_r^2\tau^{rr} & -\gamma^2v_r\tau^{rr} &  0                   &  0               \\
      \gamma^2v_r\tau^{rr}   & -\gamma^2\tau^{rr}    &  0                   &  0               \\
      0                      &  0                    & -\tau^{\theta\theta} &  0               \\
      0                      &  0                    &  0                   & -\tau^{\phi\phi} \\
    \end{array}
  \right).
\end{aligned}
\end{equation}
As $\pi^{\mu\nu}$ is traceless, {\it i.e.} $\pi^\mu_{~\mu}=0$, so $\tau^{rr}+\tau^{\theta\theta}+\tau^{\phi\phi}=0$.\\

\noindent{\bf 3. Third term of Eq.\,(\ref{2ndMIS-1})}\\

For the third term of Eq.\,(\ref{2ndMIS-1}), we have
\begin{equation}
\begin{aligned}
\nabla^{<\mu}u^{\nu>}
&\equiv \nabla^{(\mu}u^{\nu)}-\frac{1}{3}\Delta^{\mu\nu}d_\lambda u^{\lambda}\\
&=\frac{1}{2}(\nabla^\mu u^\nu+\nabla^\nu u^\mu)-\frac{1}{3}\Delta^{\mu\nu}\Theta,
\end{aligned}
\end{equation}
and
\begin{equation}
\nabla^{\mu}u^{\alpha}
=\Delta^{\mu\nu}d_\nu u^{\alpha}=\Delta^{\mu\nu}\left(\partial_\nu u^\alpha+\Gamma_{\nu\lambda}^\alpha u^\lambda\right),
\end{equation}
where $\Theta=d_\lambda u^\lambda$, and we further define $\tilde\theta \,{\equiv}\,\partial_\lambda u^\lambda$.  Then, we have
\begin{equation}
\nabla^{\mu}u^{\alpha}
=\left(
   \begin{array}{cccc}
     \nabla^{t}u^{t} & \nabla^{t}u^{r}           & 0                     \\
     \nabla^{r}u^{t} & \nabla^{r}u^{r}           & 0                     \\
     0 & 0           & \nabla^{\theta}u^{\theta} & 0                     \\
     0 & 0           & 0                         & \nabla^{\phi}u^{\phi} \\
   \end{array}
 \right)
=\left(
   \begin{array}{cccc}
    -\gamma^2v_r^2\tilde\theta & -\gamma^2v_r\tilde\theta & 0                        & 0                                    \\
    -\gamma^2v_r\tilde\theta   & -\gamma^2\tilde\theta    & 0                        & 0                                    \\
     0                         & 0                       & -\frac{1}{r^3}\gamma v_r & 0                                    \\
     0                         & 0                       & 0                        & -\frac{1}{r^3\sin^2\theta}\gamma v_r \\
   \end{array}
 \right),
\end{equation}
\begin{equation}
\begin{aligned}
&\nabla^{\langle\mu}u^{\nu\rangle}
=\left(
   \begin{array}{cccc}
   \nabla^{\langle t}u^{t\rangle} & \nabla^{\langle t}u^{r\rangle} & 0                                       & 0                                   \\
   \nabla^{\langle r}u^{t\rangle} & \nabla^{\langle r}u^{r\rangle} & 0                                       & 0                                   \\
   0                              & 0                              & \nabla^{\langle\theta}u^{\theta\rangle} & 0                                   \\
   0                              & 0                              & 0                                       & \nabla^{\langle\phi}u^{\phi\rangle} \\
   \end{array}
 \right)
=\left(
   \begin{array}{cccc}
   \gamma^2v_r^2\sigma^{rr} & \gamma^2v_r\sigma^{rr} & 0                                  & 0                                          \\
   \gamma^2v_r\sigma^{rr}   & \gamma^2\sigma^{rr}    & 0                                  & 0                                          \\
   0                        & 0                      & \frac{1}{r^2}\sigma^{\theta\theta} & 0                                          \\
   0                        & 0                      & 0                                  & \frac{1}{r^2\sin^2\theta}\sigma^{\phi\phi} \\
   \end{array}
 \right),
\end{aligned}
\end{equation}
where
\begin{equation}
\begin{aligned}
\left\{
\begin{aligned}
&\sigma^{tt}=\sigma^{tr}=\sigma^{rt}=\sigma^{rr}=\left[-\theta+\frac{1}{3}
\Theta\right]\\
&\sigma^{\theta\theta}=\sigma^{\phi\phi}=\left[-\gamma\frac{v_r}{r}+\frac{1}{3}
\Theta\right]
\end{aligned}
\right.
~\Rightarrow~
\sigma^{rr}=-2\sigma^{\theta\theta}=-2\sigma^{\phi\phi}.
\end{aligned}
\end{equation}\\

\noindent{\bf 4. Last term of Eq.\,(\ref{2ndMIS-1})}\\

The last term of Eq.\,(\ref{2ndMIS-1}) can be written as
\begin{equation}
\frac{1}{2}\pi^{\mu\nu}\tilde{\eta}Td_\lambda\left({\frac{\tau_\pi u^\lambda}{\tilde{\eta}T}}\right)
=\frac{1}{2}\pi^{\mu\nu}\tilde{\eta}T
 \left[\frac{\tau_\pi}{\tilde{\eta}T}d_\lambda u^\lambda
       +u^\lambda d_\lambda\left(\frac{\tau_\pi}{\tilde{\eta}T}\right)\right]
=\frac{1}{2}\pi^{\mu\nu}\tilde{\eta}T
 \left[\frac{\tau_\pi}{\tilde{\eta}T}\Theta+D\left(\frac{\tau_\pi}{\tilde{\eta}T}
\right)\right].
\end{equation}
Now, one can obtain the shear-viscosity related differential equations in Eq. (\ref{single-evol}) from the above derivations. \\

\end{subequations}

\end{small}

\vspace{-1mm}
\centerline{\rule{80mm}{0.1pt}}
\vspace{2mm}

\begin{multicols}{2}

\end{multicols}

\clearpage
\end{CJK*}
\end{document}